\documentclass{article}
\usepackage{graphicx} 
\usepackage{main} 
\usepackage{amsmath,amsthm,amssymb,ifthen,mathrsfs,amsfonts,bm}
\usepackage{empheq}
\usepackage{mathtools}
\usepackage{mathabx}
\usepackage{subcaption}
\usepackage{accents}
\usepackage{enumitem}
\usepackage{bigints}
\usepackage{caption}
\usepackage[capposition=below]{floatrow}
\usepackage{verbatim}

\def\BR{\mathbb{R}}

\makeatletter
\newcommand{\rn}[1]{\romannumeral #1}
\newcommand{\Rn}[1]{\expandafter\@slowromancap\romannumeral #1@}
\makeatother

\title{Trading off Relevance and Revenue in the Jobs Marketplace: Estimation, Optimization and Auction Design }
\author{Farzad Pourbabaee, Sophie Yanying Sheng, Peter McCrory, Luke Simon, Di Mo
  \textit{LinkedIn Corporation}\\
    \textit{Sunnyvale, California, USA}
    }
\date{November 2024}

\begin{document}

\maketitle
\section{Abstract}
We study the problem of position allocation in job marketplaces, where the platform determines the ranking of the jobs for each seeker. The design of ranking mechanisms is critical to marketplace efficiency, as it influences both short-term revenue from promoted job placements and long-term health through sustained seeker engagement. Our analysis focuses on the tradeoff between revenue and relevance, as well as the innovations in job auction design. We demonstrated two ways to improve relevance with minimal impact on revenue: incorporating the seekers preferences and applying position-aware auctions.

\section{1. Introduction}
In a jobs marketplace, when a seeker submits a job search request, a query of relevant jobs will be retrieved and shown to them according to a particular order. In economic terms, the posters of these jobs are the bidders and the list of all available positions/slots on the seeker’s page constitutes the goods. Since there is more than one position, the auction effectively involves multiple goods, each potentially having different values for both the bidders and the seeker.

Prevalent ranking mechanisms employ an ``auction score'' (see, e.g. \cite{ma2022online}) that integrates a monetization component from the bidders' side with a relevance component from the seekers' side. Optimizing solely for revenue prioritizes maximizing the monetary component, often leading to less relevant jobs being ranked higher due to their higher submitted bids --- a phenomenon known as bid dominance. On the other hand, optimizing for match quality focuses exclusively on the relevance component, potentially sacrificing revenue. This introduces a fundamental tension between the short-term gains of maximizing revenue and the long-term benefits of the users' satisfaction in any ranking problem. Jobs with fewer suitable candidates are naturally incentivized to allocate higher budgets to secure visibility, whereas jobs with broader appeal face less pressure to do so.

At a high level, the tension between revenue and relevance involves balancing short-term immediate revenue generation with building long-term gains through improved user engagement and satisfaction. An optimal job matching auction should achieve this balance by adopting a dynamic approach that considers both short- and long-term objectives.

Formally, when a seeker $i \in I$ submits a job search request, a query of $j \in [n]$ relevant jobs is retrieved, ranked, and displayed to them.\footnote{We adopt the convention that $[n] \coloneq \{1, \ldots, n\}$, representing the set of $n$ available jobs.} To achieve this, we construct an auction score that incorporates features of both the seeker and the job posters.

A common approach to produce the auction score is to modulate the bid value submitted by job $j$, denoted by $b_{ij}$, with the estimated click-through probability $\bar \pi_{ij}$:
\begin{equation*}
    \bar s_{ij} = \mathsf{S} (b_{ij} \cdot \bar\pi_{ij}) \,.
\end{equation*}
Under the Generalized-First-Price (GFP) auction, jobs are ranked based on their auction score, that is if $\bar s_{ij_1} \geq \bar s_{i j_2} \geq \ldots \geq \bar s_{i j_n}$, then the posting with index $j_k$ will sit in the $k$-th place.

While this mechanism is effective due to its simplicity, it is not optimal for two reasons: (\rn{1}) There is considerable dispersion in discounted-cumulative-Gain (DCG) between seekers due to the inherent sparsity in labor markets. A significant portion of (local) labor markets consists of a small number of hirers and seekers, highlighting the need to capture the heterogeneity in seekers' experience within the auction score; (\rn{2}) Sequential ranking based on the position-unaware scores overlooks the fact that the click-through rate of the same job for the same seeker varies depending on its position.

To address the first drawback, we augment the auction score with a seeker component that captures the relevance of job $j$ to seeker $i$, thereby incorporating the seeker's preferences. This enhancement is detailed in Section~2. To mitigate the second drawback, we introduce \textit{position-aware} auction scores. Specifically, in addition to the seeker index $i$ and the job index $j\in [n]$, the auction score will also depend on the position $k\in [n]$ in which the job is displayed to the seeker (see Section~3). These types of position-aware auctions are widely used in the advertising environment, where ad placement and performance depend heavily on the position within search results or web pages \cite{varian2007position, hummel2016position, colini2020ad}. 

\section{2. Relevance-Revenue Tradeoff}  
In this section, we study how we can augment the score function to account for the relevance of the job to the seeker. Formally, we introduce $\bar{\mu}_{ij}$, a model-based estimation of the relevance of job $j$ to seeker $i$.  We refer to this variable as \textsf{eRelevance}. It represents the quality of the match from the seeker's perspective, capturing how well the job aligns with their preferences and qualifications. The monetary gain of adding one more unit of relevance to seeker $i$ is denoted by $w_i$, and referred to as the \textsf{Seeker-Weight}. Hence, the augmented auction score now features two components:
\begin{equation*}
    \bar s_{ij} = \mathsf{S}\left(
        \underbrace{b_{ij} \cdot \bar\pi_{ij}}_{\text{poster component}}, 
        \underbrace{w_i  \cdot \bar\mu_{ij}}_{\text{seeker component}}
    \right)\,.
\end{equation*}
In the following, we present two approaches to calibrate the Seeker-Weight. In Section~2.1, we leverage experimental data to estimate how the relevance measurement responds to variations in the Seeker-Weight\footnote{Relevance measurement refers to well-established metrics for evaluating ranking quality in recommender systems, such as Normalized Discounted Cumulative Gain (NDCG), Mean Average Precision (MAP), and Mean Reciprocal Rank (MRR).}, thereby quantifying its causal impact. In Section~2.2, we propose a dynamic framework to determine the optimal level of the Seeker-Weight that maximizes long-term gains. Overall, the first approach focuses on causal estimation, while the second approach adopts an optimization-based strategy.

\subsection{2.1 Causal Estimation} 
A higher Seeker-Weight increases the influence of eRelevance on the final auction score, resulting in rankings that better reflect the seeker's preferences. This improved alignment boosts seeker engagement, which ultimately encourages more active participation from both seekers and posters.

A good starting point is to increase Seeker-Weight for segments of the job market for whom relevance is undesirably low. A question arises: By how much should we increase the Seeker-Weight? To answer this question we need two things: a target relevance level and a parsimonious model that maps Seeker-Weight to a relevance metric. 

\paragraph{Parsimonious Model:} Consider the following model of relevance measurement as a function of the Seeker-Weight for a particular seeker segment $g$ under the conditions that $\alpha \in (0,1)$:
\begin{equation*}
    \text{Rel}_g = \mathsf{F}_g(\text{Seeker-Weight}) = z_g \times  \text{Seeker-Weight}^{\alpha}
\end{equation*}
The parameter $z_g$ represents segment-specific efficiency in converting Seeker-Weight to relevance. Segments with higher $z_g$ achieve higher relevance levels for any given Seeker-Weight, reflecting heterogeneity in baseline relevance efficiency. The parameter $\alpha \in (0,1)$ captures the elasticity of relevance with respect to the Seeker-Weight, featuring diminishing marginal returns in terms of relevance per unit increase in the Seeker-Weight. 

We then used an experimental setting to capture the causal effect of the Seeker-Weight on the relevance metric. If the above model is a good approximation to how the marketplace works, then the observed lift in relevance across seeker segments caused by a higher Seeker-Weight should be unrelated to the baseline level of relevance (as measured in the control group). Moreover, the lift in relevance should only be a function of the change in Seeker-Weight (which is observed) and the unobserved elasticity term $\alpha$. Indeed, this is exactly what we find: Among seeker segments where the Seeker-Weight was increased from a \textit{moderate} to a \textit{high} level, relevance rose significantly on average. Conversely, among seeker segments where the weight was decreased from a \textit{moderate} to a \textit{low} level, relevance fell by a significant level.
\begin{figure}[ht]
    \centering
    \includegraphics[width=\linewidth]{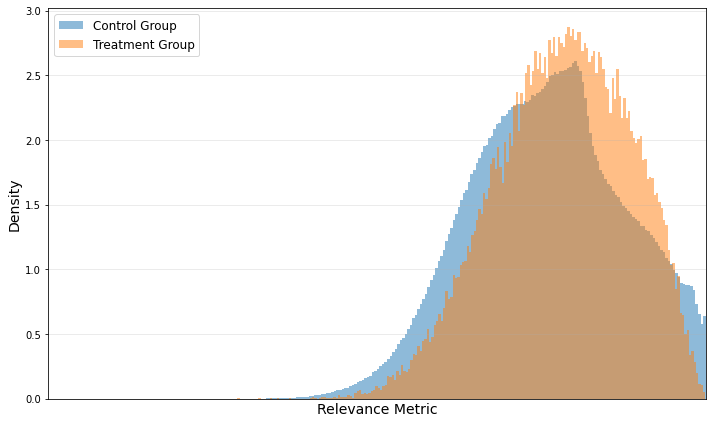} 
    \caption{Dispersion in relevance metric in experiment}
    \label{fig:elasticity}
\end{figure}

We calibrate the parsimonious model against experiment data to find elasticity of relevance with respect to the Seeker-Weight, namely, $\alpha$. We use this $\alpha$ to tune personalized Seeker-Weights in jobs search auctions, and A/B test the new weights. Figure \ref{fig:elasticity} shows that there is reduced relevance metric dispersion under new Seeker-Weights, as well as elevated relevance level for the mean and the median seeker. This improvement in relevance metric does not come at a cost of short-term revenue. 


\subsection{2.2 Dynamic Optimization}
Having pinned down the causal impact of Seeker-Weight on the seeker's relevance,
the platform's problem is now to choose an optimal Seeker-Weight that maximizes long-run discounted ``gains''. Lower weights on relevance increase immediate gains but potentially degrade future gains by reducing seeker engagement and poster participation. 
We model this dynamic problem by introducing a time dimension to the Seeker-Weight, i.e., $w_i(t)$, and assume a discount factor $\delta < 1$ across time periods. In each period $t$, the platform observes the submitted bids $b_{ij}(t)$'s, eRelevances $\bar \mu_{ij}(t)$'s, and sets the personalized Seeker-Weight $w_i(t)$. These jointly determine the auction score, which feeds into a particular mechanism of the slot assignment, and result in current gain. The platform's objective is to maximize the expected discounted gains:
\begin{equation*}
    \max_{\{w_i(t)\}} \mathbb{E}\left[\sum_{t=0}^{\infty} \delta^t \, \text{Gain}(t)\right]\,.
\end{equation*}
The dynamics are governed by a stochastic kernel $Q$ that maps the current history of the states and Seeker-Weights to the distribution over future states. Formally, the state vector $x_i(t)$ includes current bids and eRelevances, i.e.,  $x_i(t) = \{b_{ij}(t), \bar\mu_{ij}(t): j \in [n]\}$. The history $h_i(t)$ includes all past states, that is $h_i(t)=\{x_i(\tau): \tau \leq t\}$. Suppose $\sigma$ is a policy function that maps the current history $h_i(t)$ to the imputed Seeker-Weight. We can write the value of any such policy as:
\begin{equation*}
\begin{gathered}
    V_{\sigma}(h_i(t)) = \text{Gain}\left(x_i(t);\sigma(h_i(t))\right) + \\
    \delta \mathbb{E}^Q[V_{\sigma}(h_i(t+1))|h_i(t),\sigma(h_i(t))]\,.
\end{gathered}
\end{equation*}
However, this formulation faces two key challenges. First, the transition kernel $Q$ is unknown and must be learned. Second, the increasing length of the history makes optimization over the space of all policy functions computationally infeasible. To address the second challenge, we impose a Markovian structure: future states depend only on the current state and Seeker-Weight, rather than the entire history. This simplification enables us to focus on Markov strategies that map current states directly to weights. The simplified Bellman equation becomes:
\begin{equation}
\begin{gathered}
    V(x_i(t)) = \max_{\sigma} \{\text{Gain}\left(x_i(t);\sigma(x_i(t))\right) + \\
    \delta \mathbb{E}^Q[V(x_i(t+1))|x_i(t),\sigma(x_i(t))]\}\,.
\end{gathered}
\end{equation}
We solve this recursively through value function iteration, starting with $V^{(0)}(\cdot) \equiv 0$ and updating according to:

\begin{equation*}
    \begin{aligned}
        \sigma^{(m+1)}(x) &= \text{argmax}_{\sigma(x)} \Big\{\text{Gain}(x;\sigma(x)) \\
        &\quad + \delta \mathbb{E}^Q[V^{(m)}(x')|x,\sigma(x)]\Big\},
    \end{aligned}
\end{equation*}
where $m$ is the index of the iteration.

Standard dynamic programming guarantees convergence to optimal values and policies as $m \to \infty$, provided $\delta < 1$. The key empirical challenge in implementing this framework is that the transition kernel $Q$ must be learned from data. We address this through reinforcement learning methods that adaptively update beliefs about $Q$ while optimizing the Seeker-Weights. 
\section{3. Auction Design} 
\label{sec:sec3}
In the jobs marketplace, when a seeker submits a search request, a query of $n$ relevant jobs is retrieved and shown according to a particular order. In economic terms, the posters of these jobs are the ``bidders'', and the list of all available positions on the seeker’s page constitutes the ``goods''. Since there are more than one position, we are effectively selling multiple goods, with potentially different values to the bidders, as well as the seeker. That means we are in the world of \emph{multiple/heterogeneous-goods} auction design. Similar to every other mechanism design problem, here we need to determine two functions, that jointly define the overall mechanism:

\textit{Allocation rule}: also known as permutation, that prescribes the order of the postings shown to the seeker.

\textit{Transfer rule}: also known as pricing, prescribes the price per slot that is charged to each bidder.

Suppose we want to assign multiple slots at the same time to a group of bidders. The well-developed theory of selling a single item to a group of buyers \cite{krishna2009auction, myerson1981optimal} is not directly applicable to the case of allocating multiple goods. A common feature of single-good auctions is that the item should be allocated to the buyer who values it the most — the allocation is greedy. This allocation is both revenue maximizing for the seller and welfare maximizing for the buyers. However, this intuition does not immediately carry over to multiple goods, mainly because of two reasons: capacity constraints (agents cannot own multiple goods), and the possibility of bundling. We highlight these issues in the following examples, where there are two buyers Alice and Bob, and two items $x$, $y$ to be allocated.

\textbf{Example 1} (Capacity constraints): Let $v_a=(0, 1)$ and $v_b= (2, 3)$ be the valuations of Alice and Bob respectively for goods $x$ and $y$. If we allocate each good to the buyer who values it the most, then both items should be sold to Bob. In case that he has capacity constraints, then selling $y$ to Bob and $x$ to Alice is the optimal plan. Here is an example where allocating each good to the buyer who values it the most fails.

\textbf{Example 2} (Bundling): We add a third coordinate into the buyers’ valuation vectors that represents their joint valuation of getting the bundle $x$ and $y$. Let $v_a=(2, 1, 7)$ and $v_b= (0, 3, 4)$. Allocating each good to the person who values it the most means that item $x$ should be sold to Alice, and $y$ should go to Bob, yielding the overall surplus of $2 + 3 =5$. But if we sell both items to Alice the net surplus would be $7$. Thus, once again greedily allocating each object according to the \emph{maximum-demand} schedule (or first-price principle) fails.

To account for the valuation heterogeneity of different slots, we adopt position-aware scores. Specifically, a job poster $j \in [n]$ submits a bid value $b_{ij}$ for the available slots $k \in [n]$ in the search results of seeker $i \in I$. The click-through probabilities are position-aware, denoted as $\pi_{ijk}$, representing the likelihood that seeker $i$ clicks on job $j$ displayed in position $k$. Additionally, the relevance of job $j$ in position $k$ for seeker $i$ is captured by the position-aware eRelevance, $\mu_{ijk}$. These position-aware parameters can be used to generate position-aware auction scores:
\begin{equation*}
    s_{ijk} = \mathsf{S}\left(b_{ij} \cdot \pi_{ijk} ,\, w_i \cdot \mu_{ijk}\right)\,.
\end{equation*}

Below, in Section~3.1, we briefly review the application of GFP, which is inherently a position-unaware assignment method, to the above scores. Then, in Section~3.2, we extend our analysis to the assignment problem with position-aware scores, exploring the full-fledged bipartite matching approach, also known as VCG. Lastly, in Section~3.3 we present some simulations to compare these two assignment algorithms.

\subsection{3.1 Position-Unaware and Generalized First-Price Ranking}
Suppose instead of position-aware scores $\{\pi_{ijk}, \mu_{ijk}, s_{ijk}\}$, we only have their position-unaware estimates denoted by $\{\bar \pi_{ij}, \bar \mu_{ij}, \bar s_{ij}\}$, which can be found by taking the average over all slots $k\in [n]$. GFP prescription is to rank jobs based on their slot-averaged scores. Specifically, if $\bar{s}_{ij_1} \geq \bar{s}_{ij_2} \geq \ldots \geq \bar{s}_{ij_n}$, then the job with index $j_k$ is assigned to the $k$-th position. The running time is typical of any sorting algorithm, $\mathcal{O}(n \log n)$. As before, the transfer rule typically follows a ``pay-for-performance model,'' where performance is defined by metrics such as impressions, click-throughs, or job applications, which ultimately determine the revenue.

\subsection{3.2 Position-Aware and Bipartite Matching}
In this section, we study how the jobs should be ranked on the seeker’s query when we have position-aware scores as input. Given that the seeker is fixed, and to reduce clutter, we drop the seeker index $i$ from all forthcoming variables, but it should be understood from the context that all of them depend on the index of the seeker.

Recall that $j \in [n]$ refers to the job posting identifier, and there are $n$ available positions indexed by $k \in [n]$. A matching function (or an allocation rule) is represented by an one-to-one function $M: [n] \to [n]$, mapping $n$ job postings to $n$ available slots. In particular, $M(j) = k$ means that posting $j$ will take the $k$-th position on the list, and vice-versa $M^{-1}(k) = j$, means that the position $k$ is filled by the job posting $j$.

Our goal is to find a matching $M$ that maximizes the sum of (position-aware) auction scores, that is, to solve the following combinatorial maximization problem:
\begin{equation*}
    \max_{M} \sum_{j \in [n]} s_{j M(j)}\,.
\end{equation*}
This problem falls under the classic bipartite matching for which various computational algorithms were proposed. Hungarian approach (a.k.a Kuhn-Munkres algorithm \cite{kuhn1955hungarian}) provides an exact solution in $\mathcal{O}(n^3)$ time, and $\varepsilon$-approximate algorithms are known to achieve lower computational complexity \cite{bertsekas1998network, bertsekas2024new}. The theory of selling heterogeneous goods is first studied in economics literature by \cite{demange1986multi}. The authors showed that repeated application of the $\varepsilon$-approximate algorithm, by gradually sending $\varepsilon\to 0$, results in the implementation of the optimal matching function with the \textit{minimum Walrasian prices}, known as VCG transfer rule \cite{vickrey1961counterspeculation, clarke1971multipart}. More specifically, their algorithm not only outputs the optimal matching but also, as a byproduct, determines the prices for each item in the assignment problem. We only use the allocation rule of VCG (that is the same as the Hungarian output) to determine the optimal position-aware permutation of jobs, and similar to GFP, we use a pay-for-performance model to charge the bidders.

\subsection{3.3 Simulations}
In this section, we present simulations to compare the performance of GFP and VCG rankings based on two criteria: revenue and relevance. Specifically, we repeat the simulation over a fixed number of seekers $|I|$, and then take the sample average. Below, we explain the simulation steps for each $i \in I$:
\begin{enumerate}[label = (\roman*)]
    \item Draw $n$ random samples from the underlying distribution of bid values: $\{b_{i,j}: j \in [n]\}$.
    \item Draw $n^2$ position-aware random samples from the underlying distribution of pCTR: $\left\{\pi_{ijk}: j \in [n], k\in [n]\right\}$.
    \item Draw a Seeker-Weight $w_i \in \BR_+$.
    \item Draw $n^2$ position-aware random samples from the underlying distribution of eRelevance: $\left\{\mu_{ijk}: j \in [n], k\in [n]\right\}$.
    \item Compute the position aware scores $s_{ijk} = \mathsf{S}\left(b_{ij} \cdot \pi_{ijk} , w_i \cdot \mu_{ijk}\right)$. 
    \item Compute the average values over $n$ slots: $\bar \pi_{ij}$, $\bar \mu_{ij}$, and $\bar{s}_{ij}$.
    \item Find the GFP ranking based on the average scores, and compute its resultant revenue $\textsf{Rev-GFP}_i$ and relevance $\textsf{Rel-GFP}_i$.
    \item Find the VCG ranking based on the position-aware scores, and compute its resultant revenue $\textsf{Rev-VCG}_i$ and relevance $\textsf{Rel-VCG}_i$.
\end{enumerate}
Once we have the revenue and relevance per seeker $i$, we take their averages over the population of seekers $I$ and compare them across the two matching algorithms (GFP and VCG):
\begin{equation*}
    \begin{gathered}
        \textsf{Rev-GFP} \coloneq \frac{1}{|I|}\sum_{i \in I} \textsf{Rev-GFP}_i \,, \,\textsf{Rev-VCG} \coloneq \frac{1}{|I|}\sum_{i \in I} \textsf{Rev-VCG}_i \\
        \textsf{Rel-GFP} \coloneq \frac{1}{|I|}\sum_{i \in I} \textsf{Rel-GFP}_i \,, \,\textsf{Rel-VCG} \coloneq \frac{1}{|I|}\sum_{i \in I} \textsf{Rel-VCG}_i
    \end{gathered}
\end{equation*}
\begin{figure}[ht]
    \centering
    \includegraphics[width=\linewidth]{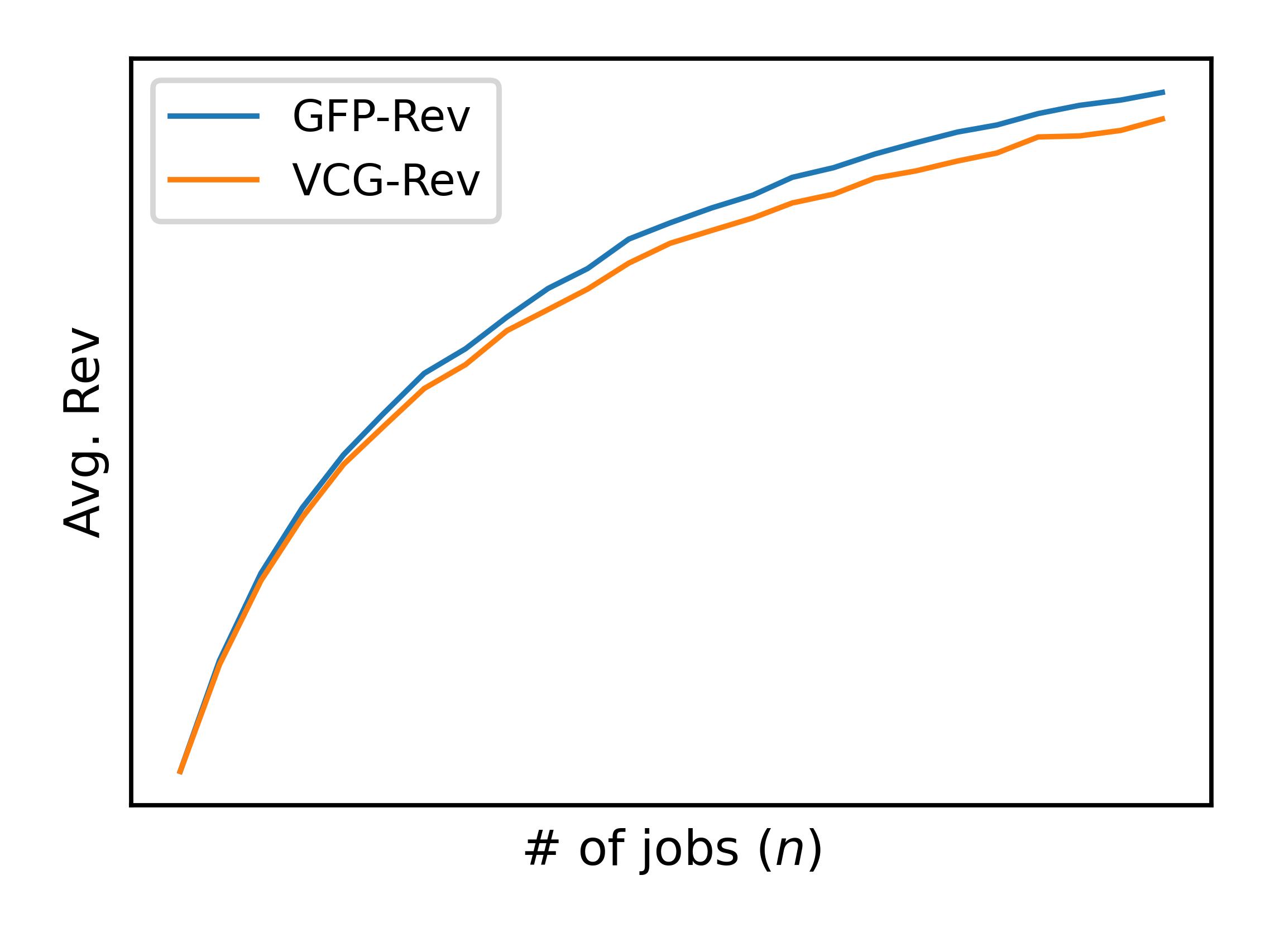} 
    \caption{Average revenue per seeker vs. \# of jobs}
    \label{fig:rev}
\end{figure}
\begin{figure}[ht]
    \centering
    \includegraphics[width=\linewidth]{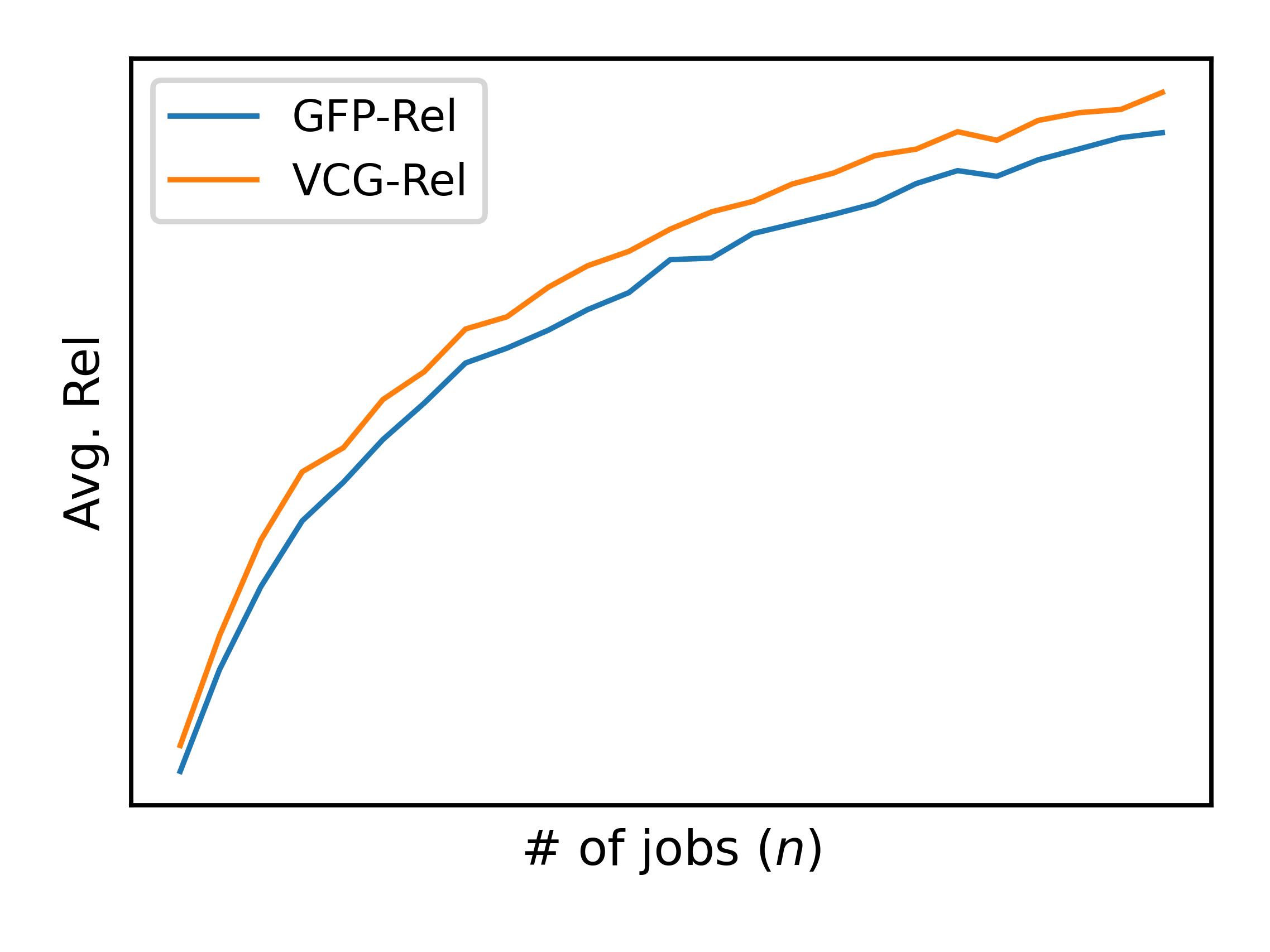} 
    \caption{Average relevance per seeker vs. \# of jobs}
    \label{fig:rel}
\end{figure}

In Figure~\ref{fig:rev}, the average revenue for both assignment algorithms are plotted as a function of the auction depth, i.e., the number of bidders or participating jobs. Both graphs show an increasing trend with $n$, which aligns with the intuition that more bidders lead to greater competition and, consequently, higher revenue. Moreover, GFP consistently extracts higher average revenue than VCG. In Figure~\ref{fig:rel}, the average relevance of both algorithms are plotted as a function of the number of participating jobs. VCG, by design, ranks jobs based on position-aware scores, whereas GFP relies on average auction scores across all positions. As a result, VCG outperforms GFP in terms of relevance.

These two graphs underscore the tradeoff between revenue and relevance. Despite incorporating seeker eRelevance into the auction score, the greedy nature of GFP continues to extract more revenue than the position-sensitive approach of VCG. However, on the relevance front, VCG demonstrates superior performance, highlighting its strength in aligning with seeker preferences.

\section{4. Concluding Remarks}
This paper makes two contributions to the design of job marketplace auctions. First, we capture the heterogeneity inherent in the jobs marketplace that leads to a wide dispersion of seekers' experiences under a uniform auction score formulation. We address this dispersion by augmenting the auction score with a parameter that allows for seeker personalization. We calibrate the parameter using (\rn{1}) a parsimonious model of relevance as a function of the new parameter, and (\rn{2}) a reinforcement learning framework to optimally balance the tension between short- and long-term gains, incorporating dynamic behavioral responses of the seekers and posters into the optimization problem. Second, as an alternative to GFP, we explore a position-aware auction for multiple slots, demonstrating improvements in relevance even without altering the auction score formulation. We expect our approach to be widely applicable in other two-sided marketplaces where similar tensions between immediate monetization and long-term platform health exist.


\bibliographystyle{plain}
\bibliography{main}
\end{document}